\documentclass[12]{iopartred}
\usepackage{amsbsy,latexsym,amsmathred}
\usepackage{amsfonts}
\usepackage{amssymb}
\usepackage[mathscr]{eucal}
\usepackage{graphicx,txfonts}
\usepackage{hyperref}

\renewcommand{\tr}{{\rm tr}\,}
\newcommand{\trnorm}[1]{\parallel #1 \parallel_1}

\newcommand{\ket}[1]{\left|{#1}\right\rangle}
\newcommand{\bra}[1]{\left\langle{#1}\right|}
\newcommand{\braket}[2]{\langle{#1}|{#2}\rangle}
\newcommand{\ketbrad}[1]{\left|{#1}\rangle\!\langle{#1}\right|}
\newcommand{\ketbra}[2]{\left|{#1}\rangle\!\langle{#2}\right|}

\newcommand\xyZ[3]{\mbox{\tiny $\kern-.3em(\!#1\kern-.2em #2\!)\kern-.14em #3\!$}}
\newcommand\Xyz[3]{\mbox{\tiny $\kern-.3em #1\kern-.14em(\! #2\kern-.14em #3\kern-.14em)\!$}}

\renewcommand{\vec}[1]{\mbox{\boldmath$#1$}}

\begin{document}
\title[Quantum learning of coherent states]{{\sf \bfseries Quantum learning of coherent states}}

\author{{\sf \bfseries Gael~Sent\'{i}s$^1$,  M\u{a}d\u{a}lin Gu\c{t}\u{a}$^2$, and Gerardo Adesso$^2$}}
\address{$^1$F\'isica Te\`orica: Informaci\'o i Fen\`omens Qu\`antics, Universitat
Aut\`{o}noma de Barcelona, 08193 Bellaterra (Barcelona), Spain}
\address{$^2$School of Mathematical Sciences, The University of Nottingham, University Park, Nottingham NG7 2RD, United Kingdom}


\begin{abstract}

We develop a quantum learning scheme for binary discrimination of coherent states of light. This is a problem of technological relevance for the reading of information stored in a digital memory.
In our setting, 
a coherent light source is used to illuminate a memory cell and retrieve its encoded bit by determining the quantum state of the reflected signal.
We consider a situation where the amplitude of the states produced by the source is not fully known, but instead this information is encoded in a large training set comprising many copies of the same coherent state.
We show that an optimal global measurement, performed jointly over the signal and the training set, provides higher successful identification rates than any learning strategy based on first estimating the unknown amplitude by means of Gaussian measurements on the training set, followed by an adaptive discrimination procedure on the signal.
By considering a simplified variant of the problem, we argue that this is the case even for non-Gaussian estimation measurements.
Our results show that, even in absence of entanglement, collective quantum measurements yield an enhancement in the readout of classical information, which is particularly relevant in the operating regime of low-energy signals.
\end{abstract}

\maketitle
\title[Quantum learning of coherent states]
\smallskip
\section{Introduction}

Programmable processors are expected to automate information processing tasks, lessening human intervention by adapting their functioning according to some input program. This adjustment, that is, the process of extraction and assimilation of relevant information to perform efficiently some task, is often called \emph{learning}, borrowing a word most naturally linked to living beings.
Machine learning is a well-established and interdisciplinary research field, broadly fitting within the umbrella of Cybernetics, that seeks to endow machines with this sort of ability,
rendering them able to
``learn'' from past experiences, perform pattern recognition and identification in scrambled data, and ultimately self-regulate \cite{MacKay2003,Bishop2006}. Algorithms featuring learning capabilities have numerous practical applications, including speech and text recognition, image analysis, and data mining.

Whereas conventional machine learning theory implicitly assumes the training set to be made of classical data,
a more recent variation, which can be referred to as quantum machine learning,
focuses on the exploration and optimisation of training with fundamentally quantum objects.
\emph{Quantum learning}~\cite{Aimeur2006}, as an area of strong foundational and technological interest, has recently raised great attention. Particularly, the use of programmable quantum processors has been investigated to address machine learning tasks such as pattern matching~\cite{Sasaki2002}, binary classification~\cite{Guta2010a,Neven2009,Sentis2012a,Pudenz2013}, feedback-adaptive quantum measurements~\cite{Hentschel2010}, learning of unitary transformations~\cite{Bisio2010}, `probably approximately correct' learning~\cite{Servedio2004}, and unsupervised clustering~\cite{Lloyd2013L}.
Quantum learning algorithms provide not only performance improvements over some classical learning problems, but they naturally have also a wider range of applicability. Quantum learning has also strong links with quantum control theory \cite{qcontrol}, and is thus becoming an increasingly significant element of the theoretical and experimental quantum information processing toolbox.

In this paper, we investigate a quantum learning scheme for the task of discriminating between two coherent states.
Coherent states stand out for their relevance in quantum optical communication theory \cite{Glauber1963,Cahill1969,Cahill1969a}, quantum information processing implementations with light, atomic ensembles, and interfaces thereof \cite{cvbook,Grosshans2003a}, and quantum optical process tomography \cite{Lobino2008a}. Lasers are widely used in current telecommunication systems, and the transmission of information can be theoretically modelled in terms of bits encoded in the amplitude or phase modulation of a laser beam. The basic task of distinguishing two coherent states in an optimal way is thus of great importance, since lower chances of misidentification translate into higher transfer rates between the sender and the receiver.

The discrimination of coherent states has been considered, so far,  within two main approaches, namely minimum-error and unambiguous discrimination, although the former is more developed. Generally, a logical bit can be encoded in two possible coherent states $\ket{\alpha}$ and $\ket{-\alpha}$, via a phase shift, or in the states $\ket{0}$ and $\ket{2\alpha}$, via amplitude modulation. Both encoding schemes are equivalent, since one can move from one to the other by applying Weyl's displacement operator $\hat{D}(\alpha)$ 
to both states\footnote{For a single mode with annihilation and creation operators $\hat{a}$ and $\hat{a}^\dagger$  respectively, the displacement operator is $\hat{D}(\alpha) = \exp\left(\alpha \hat{a}^\dagger - \alpha^\ast \hat{a}\right)$.}. In the minimum-error approach, the theoretical minimum for the probability of error is given by the Helstrom formula for discriminating two pure states~\cite{Helstrom1976}.
A variety of implementations have been devised to achieve 
this task,
e.g., the Kennedy receiver \cite{Kennedy1973}, based on photon counting; the Dolinar receiver \cite{Dolinar1973}, a modification of the Kennedy receiver with real-time quantum feedback; and the homodyne receiver\footnote{While the latter is the simplest procedure, it does not achieve optimality. However, for weak coherent states ($|\alpha|^2 \lesssim 0.4$), it yields an error probability very close to the optimal value $P_{\rm e}$, and it is optimal among all Gaussian measurements \cite{Takeoka2008}. In fact, among the three mentioned, only the Dolinar receiver is globally optimal.}. Concerning the unambiguous approach to the discrimination problem, results include the unambiguous discrimination between two known coherent states~\cite{Chefles1998a,Banaszek1999}, and its programmable version, i.e.,
when the information about the amplitude $\alpha$ enters the discrimination device in a quantum form~\cite{Sedlak2007,Sedlak2009,Bartuskova2008}.

The goal of this paper is to explore the fundamental task of discriminating between two coherent states with minimum error, when the available information about their amplitudes is incomplete. The simplest instance of such problem is a partial knowledge situation: the discrimination between the (known) vacuum state, $\ket{0}$, and some coherent state, $\ket{\alpha}$, where the value of $\alpha$ is not provided beforehand in the classical sense, but instead encoded in a number $n$ of auxiliary modes in the state $\ket{\alpha}^{\otimes n}$. Such discrimination scheme can be cast as a learning protocol with two steps: a first training stage where the auxiliary modes (the {\it training set}) are measured to obtain an estimate of $\alpha$, followed by a discrimination measurement based on this estimate. We then investigate whether this two-step learning procedure matches the performance of the most general quantum protocol, namely a global discrimination measurement that acts jointly over the auxiliary modes and the state to be identified.

Before proceeding with the derivation of our results and in order to motivate further the problem investigated in this paper, let us define the specifics of the setting in the context of a quantum-enhanced readout of classically-stored information.


Imagine a classical memory register modelled by an array of cells, where each cell contains a reflective medium with two possible reflectivities $r_0$ and $r_1$. To read the information stored in the register, one shines light into one of the cells and analyses its reflection. The task essentially consists in discriminating the two possible states of the reflected signal, which depend on the reflectivity of the medium and thus encode the logical bit stored in the cell. In a seminal paper on \emph{quantum reading} \cite{Pirandola2011}, the author takes advantage of ancillary modes to prepare an initial entangled state between those and the signal. The reflected signal is sent together with the ancillae to a detector, where a joint discrimination measurement is performed. A purely quantum resource---entanglement---is thus introduced, enhancing the probability of a successful identification of the encoded bit\footnote{In particular, in \cite{Pirandola2011} a two-mode squeezed vacuum state is shown to outperform any classical light, in the regime of few photons and high reflectivity memories.}.
This model has been later extended to the use of error correcting codes, thus defining the notion of quantum reading capacity \cite{Pirandola2011a} also studied in the presence of various optical limitations \cite{Lupo2013}.
The idea of using nonclassical light to improve the 
performance of classical information tasks
can be traced back to precursory works on \emph{quantum illumination} \cite{Lloyd2008a,Tan2008}, where the presence of a low-reflectivity object in a bright thermal-noise bath is detected with higher accuracy when entangled light is sent to illuminate the target region. For more recent theoretical and experimental developments in optical quantum imaging, illumination and reading, including studies on the role of nonclassical correlations beyond entanglement, refer e.g.~to Refs.~\cite{Nair2011,Spedalieri2012,Tej2013,RagySciRep,GenoveseExp,ZhangExp,RagyJosaB,GiovannettiDiscr,GaussianInterPower,DiscordEmpowered,DeviceIndependentReading}.

\begin{figure}[t]
\begin{center}
\includegraphics[width=8.5cm]{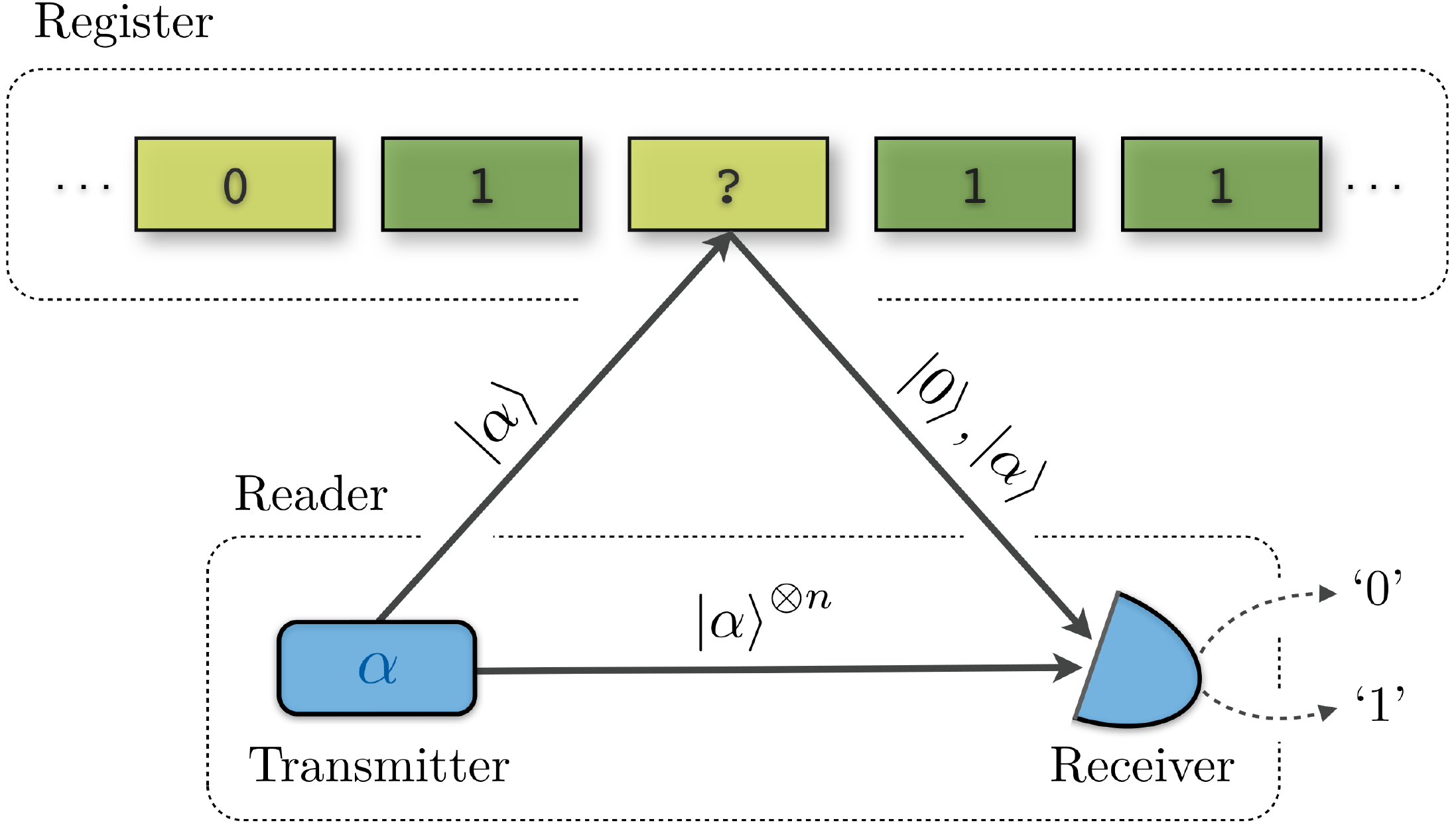}
\end{center}
\caption[Scheme of quantum reading]{A quantum reading scheme that uses a coherent signal $\ket{\alpha}$, produced by a transmitter, to illuminate a cell of a register that stores a bit of information. A receiver extracts this bit by distinguishing between the two possible states of the reflected signal, $\ket{0}$ and $\ket{\alpha}$, assisted by $n$ auxiliary modes sent directly by the transmitter.\label{ch6/fig:fig1}}
\end{figure}

In this paper we consider a reading scenario with an imperfect coherent light source and no initial entanglement involved. The proposed scheme is as follows (see Fig.~\ref{ch6/fig:fig1}). We model an ideal classical memory by a \emph{register} made of cells that contain either a transparent medium ($r_0=0$) or a highly reflective one ($r_1=1$). A \emph{reader}, comprised by a \emph{transmitter} and a \emph{receiver}, extracts the information of each cell. The transmitter is a source that produces coherent states of a certain amplitude $\alpha$. The value of $\alpha$ is not known with certainty due, for instance, to imperfections in the source, but it can be statistically localised in a Gaussian distribution around some (known) $\alpha_0$.
A signal state $\ket{\alpha}$ is sent towards a cell of the register and, if it contains the transparent medium, it goes through; if it hits the highly reflective medium, it is reflected back to the receiver in an unperturbed form. This means that we have two possibilities at the entrance of the receiver upon arrival of the signal: either nothing arrives---and we represent this situation as the vacuum state $\ket{0}$---or the reflected signal bounces back---which we denote by the same signal state $\ket{\alpha}$. To aid in the discrimination of the signal, we alleviate the effects of the uncertainty in $\alpha$ by considering that $n$ auxiliary modes are produced by the transmitter in the global state $\ket{\alpha}^{\otimes n}$ and sent directly to the receiver. The receiver then performs measurements over the signal and the auxiliary modes and outputs a binary result, corresponding with some probability to the bit stored in the irradiated cell.

We set ourselves to answer the following questions: (i) which is the optimal (unrestricted) measurement, in terms of the error probability, that the receiver can perform? and (ii) is a \emph{joint} measurement, performed over the signal together with the auxiliary modes, necessary to achieve optimality? To accomplish the set task, we first obtain the optimal minimum-error probability considering collective measurements (Section~\ref{ch6/sec:collective}).
Then, we contrast the result with that of the standard estimate-and-discriminate (E\&D) strategy, consisting in first estimating $\alpha$ by measuring the auxiliary modes, and then using the acquired information to determine the signal state by a discrimination measurement tuned to distinguish the vacuum state $\ket{0}$ from a coherent state with the estimated amplitude (Section~\ref{ch6/sec:local}).
In order to compare the performance of the two strategies, we focus on the asymptotic limit of large $n$. The natural figure of merit is the excess risk, defined as the excess asymptotic average error per discrimination when $\alpha$ is perfectly known.
We show that a collective measurement provides a lower excess risk than any Gaussian E\&D strategy, and we conjecture (and provide strong evidence) that this is the case for all local strategies (Section~\ref{ch6/sec:general}). We conclude with a summary and discussion of our results (Section~\ref{ch6/sec:final}), while some technical derivations and proofs are deferred to Appendices.
\\

\section{Collective strategy}\label{ch6/sec:collective}

The global state that arrives at the receiver can be expressed as either \mbox{$\left[ \alpha \right]^{\otimes n} \otimes [0]$} or $\left[ \alpha \right]^{\otimes n} \otimes [\alpha]$, where the shorthand notation $[\,\cdot\,]\equiv\ketbrad{\,\cdot\,}$ will be used throughout the paper.
For simplicity, we take equal \emph{a priori} probabilities of occurrence of each state. We will always consider the signal state to be that of the last mode, and all the previous modes will be the auxiliary ones.
First of all, note that the information carried by the auxiliary modes can be conveniently ``concentrated'' into a single mode by means of a sequence of unbalanced beam splitters\footnote{See, e.g., Section III A in \cite{Sedlak2008} for details.}. The action of a beam splitter over a pair of coherent states $\ket{\alpha}\otimes\ket{\beta}$ yields
\begin{equation}
\ket{\alpha}\otimes\ket{\beta} \,\longrightarrow\,
\left|\sqrt{T}\alpha+\sqrt{R}\beta\right\rangle\otimes \left|-\sqrt{R}\alpha+\sqrt{T}\beta\right\rangle \,,
\end{equation}
where $T$ is the transmissivity of the beam splitter, $R$ is its reflectivity, and $T+R=1$. A balanced beam splitter ($T=R=1/2$) acting on the first two auxiliary modes thus returns $\ket{\alpha}\otimes\ket{\alpha}\longrightarrow|\sqrt{2}\alpha\rangle\otimes\ket{0}$.
Since the beam splitter preserves the tensor product structure of the two modes, one can treat separately the first output mode and use it as input in a second beam splitter, together with the next auxiliary mode. By choosing appropriately the values of $T$ and $R$, the transformation $|\sqrt{2}\alpha\rangle\otimes\ket{\alpha}\longrightarrow|\sqrt{3}\alpha\rangle\otimes\ket{0}$ can be achieved. Applying this process sequentially over the $n$ auxiliary modes, we perform the transformation
\begin{equation}
\ket{\alpha}^{\otimes n} \,\longrightarrow\, |\sqrt{n}\alpha\rangle\otimes\ket{0}^{\otimes n-1}\,.
\end{equation}
Note that this is a deterministic process, and that no information is lost, for it is contained completely in the complex parameter $\alpha$. This operation allows us to effectively deal with only two modes. The two possible global states entering the receiver hence become $[\sqrt{n}\alpha]\otimes[0]$ and $[\sqrt{n}\alpha]\otimes[\alpha]$.

The parameter $\alpha$ is not known with certainty.
This lack of information can be embedded into the analysis by considering \emph{averaged} global states over the possible values of $\alpha$, where the choice of the prior probability distribution accounts for the prior knowledge that we might already have.
One readily sees that a flat prior distribution for $\alpha$, representing a limiting situation of complete ignorance, is not reasonable in this particular setting.
On the one hand, such prior would yield divergent average states of infinite energy, since the phase space is infinite. On the other hand, in a real situation it is not reasonable at all to assume that \emph{all} amplitudes $\alpha$ are equally probable\footnote{Nonetheless, for finite dimensional systems, assuming a uniform prior distribution can be better justified and very useful; see, e.g., \cite{Sentis2010withE}.}.
The usual procedure in these cases is to consider that a small number of auxiliary modes is used to make a rough estimation of $\alpha$, such that our prior becomes a Gaussian probability distribution centred at $\alpha_0$, whose width goes as $\sim 1/\sqrt{n}$ \footnote{Since we are interested in comparing the asymptotic performance of discrimination strategies in the limit of large $n$, the number of modes used for the rough estimation is negligible, i.e., $\tilde{n}=n^{1-\epsilon}$. Then, it can be shown that $\alpha$ belongs to a neighbourhood of size $n^{-1/2+\epsilon}$ centred at $\alpha_0$, with probability converging to one (this is shown, though in a classical statistical context, in \cite{Gill1995}). Moreover, this happens to be true for any model of i.i.d. quantum states $\rho$ (regardless their dimensionality), hence the analysis of the asymptotic behaviour of any estimation model of this sort can be restricted to a local Gaussian model, centred at a fixed state $\rho_0$. This is known as \emph{local asymptotic normality} \cite{GutaLANQS,GutaLANQubits,GutaLANQudits,Gill2013}.}.
Under these considerations, we express the true amplitude $\alpha$ as
\begin{equation}\label{ch6/alphaisalpha0}
\alpha \approx \alpha_0 + u/\sqrt{n} \, , \quad u \in \mathbb{C} \,,
\end{equation}
where the parameter $u$ follows the Gaussian distribution
\begin{equation}\label{ch6/gaussian}
G(u) = \frac{1}{\pi \mu^2} e^{-u^2/\mu^2} \,.
\end{equation}
%
To avoid divergences, we have introduced the free parameter $\mu$ as a temporal energy cut-off that defines the width of $G(u)$. After obtaining expressions for the excess risks in the asymptotic regime of large $n$, we will remove the cut-off dependence by taking the limit $\mu\to\infty$.

Exploiting the prior information acquired through the rough estimation, that is using Eqs.~\eqref{ch6/alphaisalpha0} and \eqref{ch6/gaussian}, we compute the average global states arriving at the receiver
\begin{eqnarray}
\sigma_1 &=& \int G(u) \, [ \sqrt{n} \alpha_0 + u ] \otimes [0] \,d^2u \, ,\\
\sigma_2 &=& \int G(u) \, [ \sqrt{n} \alpha_0 + u ] \otimes [\alpha_0 + u/\sqrt{n}\,]\, d^2u \, .
\end{eqnarray}
The optimal measurement to determine the state of the signal is the Helstrom measurement for the discrimination of the states $\sigma_1$ and $\sigma_2$~\cite{Helstrom1976}, that yields the average minimum-error probability\footnote{Note that, \emph{sensu stricto}, the dependence of $P_{\rm e}^{\rm opt}(n)$ on the localisation parameter $\alpha_0$ should be made explicit. Keep in mind that, in general, all quantities computed in this paper will depend on $\alpha_0$. Thus for the sake of notation clarity, we omit it hereafter when no confusion arises.}
\begin{equation}\label{ch6/perror col}
P_{\rm e}^{\rm opt}(n)= \frac{1}{2} \left(1-\frac{1}{2}\trnorm{\sigma_1-\sigma_2} \right) \,,
\end{equation}
where $\trnorm{M}=\tr \sqrt{M^\dagger M}$ denotes the trace norm of the operator $M$.
The technical difficulty in computing $P_{\rm e}^{\rm opt}(n)$ resides in the fact that $\sigma_1-\sigma_2$ is an infinite-dimensional full-rank matrix, hence its trace norm does not have a computable analytic expression for arbitrary finite $n$. Despite this, one can still resort to analytical methods in the asymptotic regime $n\to\infty$ by treating the states perturbatively.

To ease this calculation, we first apply the displacement operator
\begin{equation}\label{ch6/displacement}
\hat{D}(\alpha_0) = \hat{D}_1 (-\sqrt{n}\alpha_0) \otimes \hat{D}_2(-\alpha_0)
\end{equation}
to the states $\sigma_1$ and $\sigma_2$, where $\hat{D}_1$ ($\hat{D}_2$) acts on the first (second) mode, and we obtain the displaced global states
\begin{eqnarray}
\bar{\sigma}_1 &=& \hat{D}(\alpha_0)\sigma_1\hat{D}^\dagger(\alpha_0) = \int G(u) \left[  u \right] \otimes \left[-\alpha_0 \right] d^2u \,,\label{ch6/barsigma1}\\
\bar{\sigma}_2 &=& \hat{D}(\alpha_0)\sigma_2\hat{D}^\dagger(\alpha_0) = \int G(u) \left[  u \right] \otimes [ u/\sqrt{n}\,]\, d^2u \,.\label{ch6/barsigma2}
\end{eqnarray}
Since both states have been displaced by the same amount, the trace norm does not change, i.e., $\trnorm{\sigma_0-\sigma_1}=\trnorm{\bar{\sigma}_0-\bar{\sigma}_1}$. Eq.~\eqref{ch6/barsigma1} directly yields
\begin{equation}\label{ch6/barsigma1exp}
\bar{\sigma}_1 = \sum_{k=0}^\infty c_k [k] \otimes [-\alpha_0] \, ,
\end{equation}
where $c_k = \mu^{2k}/[(\mu^2+1)^{k+1}]$ and $\{\ket{k}\}$ is the Fock basis. Note that, as a result of the average, the first mode in Eq.~\eqref{ch6/barsigma1exp} corresponds to a thermal state with average photon number $\mu^2$. Note also that the $n$-dependence is entirely in $\bar{\sigma}_2$. In the limit $n\to\infty$, we can expand the second mode of $\bar{\sigma}_2$ 
as
\begin{equation}
|u/\sqrt{n}\,\rangle = e^{-\frac{|u|^2}{2n}} \sum_{k=0}^{\infty} \frac{(u/\sqrt{n})^k}{\sqrt{k!}} \ket{k} \,.
\end{equation}
Then, up to order $1/n$ its asymptotic expansion gives
\begin{align}\label{ch6/series_u}
[u/\sqrt{n}\,] &\simeq \ketbrad{0} + \frac{1}{\sqrt{n}} \left( u \ketbra{1}{0} + u^* \ketbra{0}{1} \right) \nonumber\\
&\!\!\!\!\!\!\!+ \frac{1}{n} \left\{ |u|^2 \left( \ketbrad{1}-\ketbrad{0}\right) + \frac{1}{\sqrt{2}} \left[ u^2 \ketbra{2}{0} + \left(u^*\right)^2 \ketbra{0}{2} \right] \right\} \, .
\end{align}
%
%
Inserting Eq.~\eqref{ch6/series_u} into Eq.~\eqref{ch6/barsigma2} and computing the corresponding averages of each term in the expansion, we obtain a state of the form
\begin{equation}\label{ch6/barsigma2exp}
\bar{\sigma}_2 \simeq \bar{\sigma}_2^{(0)} + \frac{1}{\sqrt{n}}\bar{\sigma}_2^{(1)} + \frac{1}{n}\bar{\sigma}_2^{(2)} \,.
\end{equation}
We can now use Eqs.~\eqref{ch6/barsigma1exp} and \eqref{ch6/barsigma2exp} to compute the trace norm $\trnorm{\bar{\sigma}_1-\bar{\sigma}_2}$ in the asymptotic regime of large $n$, up to order $1/n$, by applying perturbation theory. The explicit form of the terms in Eq.~\eqref{ch6/barsigma2exp}, as well as the details of the computation of the trace norm, are given in 
\ref{appD/sec:tracenormcol}.
Here we just show the result: the average minimum-error probability $P_{\rm e}^{\rm opt}(n)$, defined in Eq.~\eqref{ch6/perror col}, can be written in the asymptotic limit as
\begin{equation}\label{ch6/perror col2}
P_{\rm e}^{\rm opt} \equiv P_{\rm e}^{\rm opt}(n\to\infty) \simeq \frac{1}{2} \left[1-\sqrt{1-e^{-|\alpha_0|^2}} -\frac{1}{2n}\left(\Lambda_+^{(2)}-\Lambda_-^{(2)} \right) \right] \,,
\end{equation}
where $\Lambda_\pm^{(2)}$ is given by Eq.~\eqref{appD/Lambda2pm}.
\\

\subsection*{Excess risk}

The figure of merit that we use to assess the performance of our protocol is the \textit{excess risk}, that we have defined as the difference between the asymptotic average error probability $P_{\rm e}^{\rm opt}$ and the average error probability for the optimal strategy when $\alpha$ is perfectly known. As we said at the beginning of the section, the true value of $\alpha$ is $\alpha_0+u/\sqrt{n}$ for a particular realisation, thus knowing $u$ equates knowing $\alpha$. The minimum-error probability for the discrimination between the \emph{known} states $\ket{0}$ and $\ket{\alpha_0+u/\sqrt{n}}$, $P_{\rm e}^*(u,n)$, averaged over the Gaussian distribution $G(u)$, takes the form
\begin{align}
P_{\rm e}^*(n) &= \int G(u) \,P_{\rm e}^* (u,n) \,d^2u \nonumber\\
&= \int G(u) \,\frac{1}{2} \left(1-\sqrt{1-|\!\braket{0}{\alpha_0+u/\sqrt{n}}\!|^2} \right) d^2u \,. \label{ch6/perror known int}
\end{align}
To compute this integral we do a series expansion of the overlap in the limit $n\rightarrow \infty$ and 
integrate the resulting terms (see 
\ref{ch6/sec:gaussian integrals}).
After some algebra we obtain
\begin{equation}\label{ch6/perror known}
P_{\rm e}^* \equiv P_{\rm e}^*(n\to\infty) \simeq \frac{1}{2} \left(1-\sqrt{1-e^{-|\alpha_0|^2}} + \frac{1}{n} \Lambda^* \right)\,,
\end{equation}
where
\begin{equation}\label{ch6/perror known delta}
\Lambda^* = \frac{\mu^2 \left[ 2\left(e^{-|\alpha_0|^2}-1\right) +|\alpha_0|^2 \left(2-e^{-|\alpha_0|^2}\right) \right]}{4 \left(e^{|\alpha_0|^2}-1\right) \sqrt{1-e^{-|\alpha_0|^2}}} \,.
\end{equation}
The excess risk is then given by Eqs.~\eqref{ch6/perror col2} and \eqref{ch6/perror known} as
\begin{equation}
R^{\rm opt}_\mu = n \left(P_{\rm e}^{\rm opt} - P_{\rm e}^*\right) \,.
\end{equation}
Finally, we remove the cut-off imposed at the beginning by taking the limit $\mu \rightarrow \infty$ and we obtain
\begin{equation}\label{ch6/excess_risk_opt}
R^{\rm opt} =\lim_{\mu\to\infty} R^{\rm opt}_\mu = \frac{|\alpha_0|^2 e^{-|\alpha_0|^2/2}\left(2e^{|\alpha_0|^2}-1\right)}{16 \left( e^{|\alpha_0|^2}-1 \right)^{3/2}} \,.
\end{equation}
Note that the excess risk only depends on the module of $\alpha_0$, i.e., on the average distance between $\ket{\alpha}$ and $\ket{0}$. The excess risk is thus phase-invariant, as it should.

Eq.~\eqref{ch6/excess_risk_opt} is the first piece of information we need to address the main question posed at the beginning, namely whether the optimal performance of the collective strategy 
is achievable by an estimate-and-discriminate (E\&D) strategy.
We now move on towards the second piece.
\\

\section{Estimate \& Discriminate strategy}\label{ch6/sec:local}

An alternative---and more restrictive---strategy to determine the state of the signal consists in the natural combination of two fundamental tasks: state estimation, and state discrimination of known states. In such an 
E\&D strategy, \emph{all} auxiliary modes are used to estimate the unknown amplitude $\alpha$. Then, the obtained information is used to tune a discrimination measurement over the signal that distinguishes the vacuum state from a coherent state with the estimated amplitude. In this Section we find the optimal E\&D strategy based on Gaussian measurements and compute its excess risk $R^{\rm E\&D}$. Then, we compare the result with that of the optimal collective strategy $R^{\rm opt}$.

The most general Gaussian measurement that one can use to estimate the state of the auxiliary mode $|\sqrt{n}\alpha\rangle$ is a \emph{generalised heterodyne measurement},
represented by a positive operator-valued measure (POVM) with elements
\begin{equation}\label{ch6/localpovm}
E_{\bar{\beta}} = \frac{1}{\pi} \, [ \bar{\beta},r,\phi ] \,,
\end{equation}
i.e., projectors onto pure Gaussian states with amplitude $\bar{\beta}$ and squeezing $r$ along the direction $\phi$. The outcome of such heterodyne measurement $\bar{\beta}=\sqrt{n}\beta$ produces an estimate for $\sqrt{n}\alpha$, hence $\beta$ stands for an estimate of $\alpha$\footnote{In our notation, the outcome of the measurement also labels the estimate, so $\beta$ stands for both indistinctly. This should generate no confusion, since the trivial guess function that uses the outcome $\bar{\beta}$ to produce the estimate $\beta$ does not vary throughout the paper.}. Upon obtaining $\bar{\beta}$, the prior information that we have about $\alpha$ gets updated according to Bayes' rule, so that now the signal state can be either $[0]$ or some state $\rho(\beta)$. The form of this second hypothesis is given by
\begin{equation}
\rho(\beta) = \int p(\alpha|\beta) [\alpha] d^2\alpha \,,
\end{equation}
where $p(\alpha|\beta)$ encodes the \emph{posterior} information that we have acquired via the heterodyne measurement. It represents the conditional probability of the state of the auxiliary mode being $\ket{\sqrt{n}\alpha}$, given that we obtained the 
outcome $\bar{\beta}$. Bayes' rule dictates
\begin{equation}
p(\alpha|\beta) = \frac{p(\beta|\alpha) p(\alpha)}{p(\beta)} \,,
\end{equation}
where $p(\beta|\alpha)$ is given by (see 
\ref{appD/sec:heterodyne_prob})
\begin{equation}\label{ch6/heterodyne_prob_ab}
p(\beta|\alpha) = \frac{1}{\pi \cosh r} e^{-|\sqrt{n} \alpha - \bar{\beta}|^2-{\rm Re}[(\sqrt{n} \alpha-\bar{\beta})^2 e^{-i 2 \phi}] \tanh r} \,,
\end{equation}
$p(\alpha)$ is the prior information of $\alpha$ before the heterodyne measurement, and
\begin{equation}
p(\beta) = \int p(\alpha) p(\beta|\alpha) d^2\alpha
\end{equation}
is the total probability of giving the estimate $\beta$.

The error probability of the E\&D strategy, averaged over all possible estimates $\beta$, is then
\begin{equation}\label{ch6/perror_eyd}
P_{\rm e}^{\rm E\&D}(n) = \frac{1}{2} \left(1-\frac{1}{2} \int p(\beta) \trnorm{[0]-\rho(\beta)} d^2\beta \right) \,.
\end{equation}
Note that the estimate $\beta$ depends ultimately on the number $n$ of auxiliary modes, hence the explicit dependence in the left-hand side of Eq.~\eqref{ch6/perror_eyd}.

We are interested in the asymptotic expression of Eq.~\eqref{ch6/perror_eyd}, so let us now focus on the $n\to\infty$ scenario. Recall that an initial rough estimation of $\alpha$ permits the localisation of the prior $p(\alpha)$ around a central point $\alpha_0$, such that $\alpha \approx \alpha_0 + u/\sqrt{n}$, where $u$ is distributed according to $G(u)$, defined in Eq.~\eqref{ch6/gaussian}. Consequently, the estimate $\beta$ will also be localised around the same point, i.e., $\beta\approx\alpha_0+v/\sqrt{n}$, $v\in\mathbb{C}$. As a result, we can effectively shift from amplitudes $\alpha$ and $\beta$ to a local Gaussian model around $\alpha_0$, parameterised by $u$ and $v$. According to this new model, we make the following transformations:
\begin{eqnarray}
p(\alpha) &\rightarrow& G(u) \,,\\
p(\beta|\alpha) &\rightarrow& p(v|u) = \frac{1}{\pi \cosh r} e^{-|u-v|^2-{\rm Re}[(u-v)^2] \tanh r} \,,\\
p(\beta) &\rightarrow& p(v) = \int p(v|u) G(u) du \nonumber \\
&\,&= \frac{1}{\pi \cosh r} \frac{1}{\sqrt{1+\mu^2\left(2+\frac{\mu^2}{\cosh^2 r}\right)}} \nonumber\\
&\,&\times\; {\rm exp}\left(\frac{|v|^2\left(1+\frac{\mu^2}{\cosh^2 r}\right)+{\rm Re}[v^2]\tanh r}{\mu^4 \tanh^2 r-\left(\mu^2+1\right)^2}\right) \,,\label{ch6/pv}\\
p(\alpha|\beta) &\rightarrow& p(u|v)=\frac{p(v|u) G(u)}{p(v)} \,,\label{ch6/puv}
\end{eqnarray}
where, for simplicity, we have assumed $\alpha_0$ to be real. Note that this can be done without loss of generality. Note also that, by the symmetry of the problem, this assumption implies $\phi=0$.

The shifting to the local model 
transforms the trace norm in Eq.~\eqref{ch6/perror_eyd} as
\begin{equation}\label{ch6/tracenorm_eyd}
\trnorm{[0]-\rho(\beta)} \quad\rightarrow\quad \trnorm{[-\alpha_0]-\rho(v)} \,,
\end{equation}
where
\begin{equation}
\rho(v)=\int p(u|v) \, [ u/\sqrt{n} ] \, d^2u \,.
\end{equation}
To compute the explicit expression of $\rho(v)$ we proceed as in the collective strategy. That is, we expand $[u/\sqrt{n}]$ in the limit \mbox{$n\rightarrow\infty$} up to order $1/n$, as in Eq.~\eqref{ch6/series_u}, and we compute the trace norm using perturbation theory (see 
\ref{ch6/sec:eyd_tracenorm} for details).
The result allows us to express the asymptotic average error probability of the E\&D strategy as
\begin{equation}\label{ch6/perror_eyd_asymp}
P_{\rm e}^{\rm E\&D} \equiv P_{\rm e}^{\rm E\&D}(n\to\infty) \simeq \frac{1}{2}\left(1-\sqrt{1-e^{-\alpha_0^2}} + \frac{1}{n} \Delta^{\rm E\&D} \right)\,,
\end{equation}
where $\Delta^{\rm E\&D}$ is given by Eq.~\eqref{appD/DeltaEyD}.
\\

\subsection*{Excess risk}

The excess risk associated to the E\&D strategy is generally expressed as
\begin{equation}\label{ch6/excess_risk_eyd}
R^{\rm E\&D}(r) = n \lim_{\mu\to\infty} \left(P_{\rm e}^{\rm E\&D}-P_{\rm e}^*\right) \,,
\end{equation}
where $P_{\rm e}^*$ is the error probability for known $\alpha$, given in Eq.~\eqref{ch6/perror known}, and $P_{\rm e}^{\rm E\&D}$ is the result from the previous section, i.e., Eq.~\eqref{ch6/perror_eyd_asymp}.
The full analytical expression for $R^{\rm E\&D}(r)$ is given in Eq.~\eqref{appD/excessrisk_eyd_r}.
Note that we have to take the limit $\mu\to\infty$ in the excess risk, as we did for the collective case. Note also that all the expressions calculated so far explicitly depend on the squeezing parameter $r$ (apart from $\alpha_0$). This parameter stands for the squeezing of the generalised heterodyne measurement in Eq.~\eqref{ch6/localpovm}, which we have left unfixed on purpose. As a result, we now define, through the squeezing $r$, the optimal heterodyne measurement over the auxiliary mode to be that which yields the lowest excess risk \eqref{ch6/excess_risk_eyd}, i.e.,
\begin{equation}\label{ch6/excess_risk_eyd_2}
R^{\rm E\&D} = \min_{r} R^{\rm E\&D} (r) \,.
\end{equation}

To find the optimal $r$, we look at the parameter estimation theory of Gaussian models (see, e.g., \cite{Gill2013}). In a generic two-dimensional Gaussian shift model, the optimal measurement for the estimation of a parameter $\theta = (q,p)$ is a generalised heterodyne measurement\footnote{This is the case whenever the covariance of the Gaussian model is known, and the mean is a linear transformation of the unknown parameter.} of the type \eqref{ch6/localpovm}. Such measurement yields a quadratic risk of the form
\begin{equation}
R_{\hat{\theta}}=\int p(\theta) ((\hat{\theta}-\theta)^T G (\hat{\theta}-\theta)) d^2\theta \,,
\end{equation}
where $p(\theta)$ is some probability distribution, $\hat{\theta}$ is an estimator of $\theta$, and $G$ is a two-dimensional matrix. One can always switch to the coordinates system in which $G$ is diagonal, $G={\rm diag}(g_q,g_p)$, to write
\begin{equation}\label{ch6/quadratic_risk}
R_{\hat{\theta}}=g_q \int p(\theta) (\hat{q}-q)^2 d^2\theta + g_p \int p(\theta) (\hat{p}-p)^2 d^2\theta \,.
\end{equation}
It can be shown \cite{Gill2013} that the optimal squeezing of the estimation measurement, i.e., that for which the quadratic risk $R_{\hat{\theta}}$ is minimal, is given by
\begin{equation}
r=\frac{1}{4}\ln \left(\frac{g_q}{g_p}\right) \,.
\end{equation}
We can then simply compare Eq.~\eqref{ch6/quadratic_risk} with Eq.~\eqref{ch6/excess_risk_eyd} to deduce the values of $g_q$ and $g_p$ for our case. By doing so, we obtain that the optimal squeezing reads
\begin{figure}[t]
\begin{center}
\includegraphics[width=8.5cm]{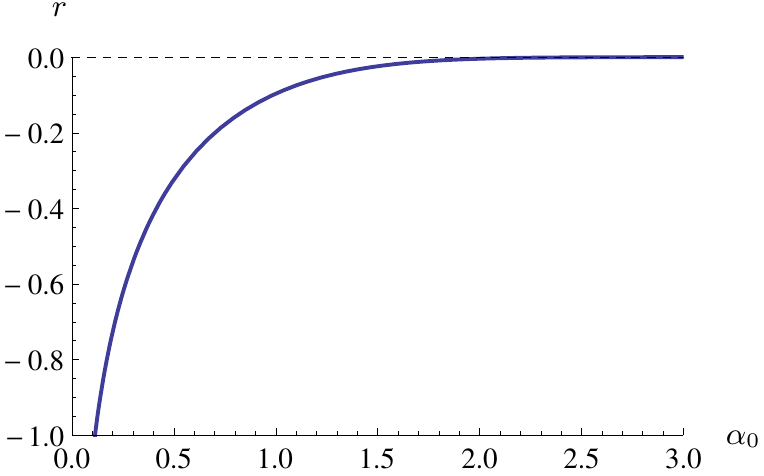}
\end{center}
\caption[Optimal squeezing for the generalised heterodyne measurement in a E\&D strategy]{Optimal squeezing $r$ for the generalised heterodyne measurement in a E\&D strategy, as a function of $\alpha_0$.\label{ch6/fig:fig2}}
\end{figure}
\begin{equation}\label{ch6/optimalsqueezing}
r = \frac{1}{4} \ln \left(\frac{f(\alpha_0)+\alpha_0^2}{f(\alpha_0)-\alpha_0^2}\right) \,,
\end{equation}
where
\begin{equation*}
f(\alpha_0)=  2 e^{\alpha_0^2} \left(e^{\alpha_0^2}-1\right)\left(\sqrt{1-e^{-\alpha_0^2}}-1\right)
+ \alpha_0^2\left(1-2e^{\alpha_0^2}\sqrt{1-e^{-\alpha_0^2}}\right) \,.
\end{equation*}%

Eq.~\eqref{ch6/optimalsqueezing} tells us that the optimal squeezing $r$ is a function of $\alpha_0$ that takes negative values, and asymptotically approaches zero when $\alpha_0$ is large (see Fig.~\ref{ch6/fig:fig2}). This means that the optimal estimation measurement over the auxiliary mode is comprised by projectors onto coherent states, antisqueezed along the line between $\alpha_0$ and the origin (which represents the vacuum) in phase space. In other words, the estimation is tailored to have better resolution along that axis because of the subsequent discrimination of the signal state. This makes sense: since the error probability in the discrimination depends primarily on the distance between the hypotheses, it is more important to estimate this distance more accurately rather than along the orthogonal direction.
For large amplitudes, the estimation converges to a (standard) heterodyne measurement with no squeezing. As $\alpha_0$ approaches 0 the states of the signal become more and more indistinguishable, and the projectors of the heterodyne measurement approach infinitely squeezed coherent states, thus converging to a homodyne measurement.

\begin{figure}[t]
\begin{center}
\includegraphics[width=8.5cm]{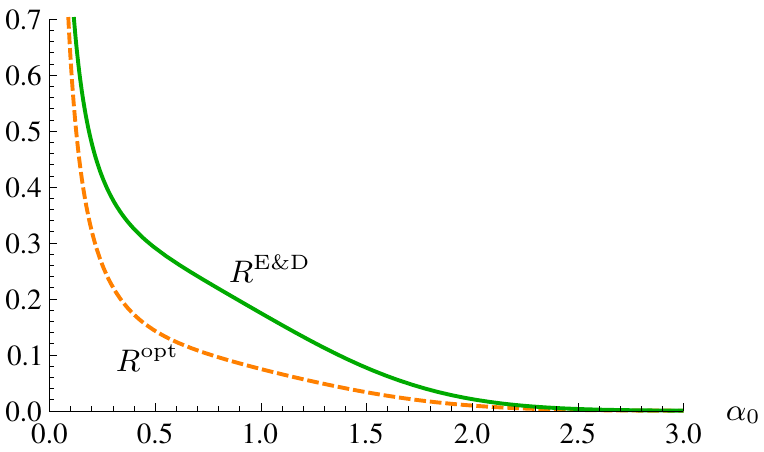}
\end{center}
\caption[Excess risk for the collective strategy, $R^{\rm opt}$, and for the E\&D strategy, $R^{\rm E\&D}$]{Excess risk for the collective strategy, $R^{\rm opt}$, and for the E\&D strategy, $R^{\rm E\&D}$, as a function of $\alpha_0$.\label{ch6/fig:fig3}}
\end{figure}

Inserting Eq.~\eqref{ch6/optimalsqueezing} into Eq.~\eqref{ch6/excess_risk_eyd_2} we finally obtain the expression of $R^{\rm E\&D}$ as a function of $\alpha_0$, which we can now compare with the excess risk for the collective strategy $R^{\rm opt}$, given in Eq.~\eqref{ch6/excess_risk_opt}. We plot both functions in Fig.~\ref{ch6/fig:fig3}. For small amplitudes, say in the range $0.3 \lesssim \alpha_0 \lesssim 1.5$, there is a noticeable difference in the performance of the two strategies, reaching more than a factor two at some points. We also observe that the gap closes for large amplitudes $\alpha_0 \rightarrow \infty$; this behaviour is expected, since the problem becomes essentially classical when the energy of the signal is sufficiently large. Interestingly, very weak amplitudes $\alpha_0 \rightarrow 0$ also render the two strategies almost equivalent.\\

\section{General estimation measurements}\label{ch6/sec:general}

We have showed that a local strategy based on the estimation of the auxiliary state via a generalised heterodyne measurement, followed by the corresponding discrimination measurement on the signal mode, performs worse than the most general (collective) strategy. However, the considered E\&D procedure does not encompass \emph{all} local strategies. The heterodyne measurement, although with some nonzero squeezing, still 
detects the phase space around $\alpha_0$ in a Gaussian way, i.e., up to second moments. In principle, one might expect that a more general measurement that produces a non-Gaussian probability distribution for the estimate $\beta$ might perform better in terms of the excess risk, and even possibly match the optimal performance, closing the gap between the curves in Fig.~\ref{ch6/fig:fig3}.
Here we show that the observed difference in performance between the collective and the local strategy is not due to lack of generality of the latter. We do so by considering a simplified although nontrivial version of the problem that allows us to obtain a fully general solution.

The intuitive reason why one could think, at first, that a non-Gaussian probability distribution for $\beta$ might give an advantage is the following.
Imagine that  $\alpha$ is further restricted to be on the positive real axis. Then, the true $\alpha$ is either to the left of $\alpha_0$ or to the right, depending on the sign of the local parameter $u$. In the former case, $\alpha$ is closer to the vacuum, so the error in discriminating between them two is larger than for the states on the other side. One would then expect that an ideal strategy should better estimate the negative values of the parameter $u$, compared to the positive ones. Gaussian measurements like the heterodyne detection do not contemplate this situation, as they are translationally invariant, and that might be the reason behind the gap in Fig.~\ref{ch6/fig:fig3}.

To test this, we design the following simple example. Since the required methods are a straightforward extension of the ones used in the previous sections, we only sketch the procedure without showing any explicit calculation.
Imagine now that the true value of $\alpha$ is not Gaussian distributed around $\alpha_0$, but it can only take the two values $\alpha=\alpha_0 \pm 1/\sqrt{n}$, representing the states that are closer to the vacuum and further away. Having only two possibilities for $\alpha$ allows us to solve analytically the most general local strategy, since estimating the auxiliary state becomes a discrimination problem between the states $\ket{\sqrt{n}\alpha_0+1}$ and $\ket{\sqrt{n}\alpha_0-1}$. The measurement that distinguishes the two possibilities is a two-outcome POVM $\mathcal{E}=\{[e_+],[e_-]\}$\footnote{Note that we have chosen the POVM elements to be rank-1 projectors. This is no loss of generality. Due to the convexity properties of the trace norm, POVMs with higher-rank elements cannot be optimal.}. We use the displacement operator~\eqref{ch6/displacement} to shift to the local model around $\alpha_0$, such that the state of the auxiliary mode is now either $\ket{1}$ or $\ket{-1}$. Note that, without loss of generality, one can confine the POVM vectors to the (Bloch) plane spanned by $\ket{1}$ and $\ket{-1}$, so that $\ket{e_+}$ and $\ket{e_-}$ are real linear combinations of these. Indeed, any component orthogonal to this plane would give no aid to distinguish the hypotheses. This allows us to express the probabilities of correctly identifying each state as
\begin{eqnarray}
p_+ &=& |\!\braket{e_+}{1}\!|^2\equiv c^2 \,, \\
 p_- &=& |\!\braket{e_-}{-1}\!|^2 = 1-\left(c\,\chi - \sqrt{1-c^2}\sqrt{1-\chi^2}\right)^2 \,,
\end{eqnarray}
where $\chi =\braket{1}{-1}=e^{-2} $, and the overlap $c$ completely parametrises the measurement $\mathcal{E}$.
If the optimal estimation measurement is indeed asymmetric,
it should happen that $p_+ < p_-$, i.e., that the probability of a correct identification is greater for the state $\ket{-1}$ than for $\ket{1}$.

From now on we proceed as for the Gaussian E\&D strategy. We first compute the posterior state of the signal mode according to Bayes' rule. Then, we compute the optimal error probability in the discrimination of $[-\alpha_0]$ and the posterior state, which is a combination of $[1/\sqrt{n}]$ and $[-1/\sqrt{n}]$, weighted by the corresponding posterior probabilities. The $c$-dependence is carried by these probabilities. Going to the asymptotic limit $n\to\infty$, applying perturbation theory to compute the trace norm, and averaging the result over the two possible outcomes in the discrimination of the signal state, we finally obtain the asymptotic average error probability for the local strategy as a function of $c$.
The asymptotic average error probability for the optimal collective strategy in this simple case is obtained exactly along the same lines as shown in Section~\ref{ch6/sec:collective}, and the one for \emph{known} states is given by the asymptotic expansion of Eq.~\eqref{ch6/perror known int}, substituting the average over $G(u)$ appropriately.

Now we can compute the excess risk for the local and collective strategy, and optimise the local one over $c$. As already advanced at the beginning, the optimal solution yields $c^*=\left(\sqrt{1+\chi}+\sqrt{1-\chi}\right)/2$, and therefore $p_+=p_-$. That is, the POVM $\mathcal{E}$ is symmetric with respect to the vectors $\ket{1}$ and $\ket{-1}$, hence both hypotheses receive the same treatment by the measurement in charge of determining the state of the auxiliary mode. Moreover, the gap between the excess risk of both strategies remains. This result leads us to conjecture that the optimal collective strategy performs better than \emph{any} local strategy.
\\

\section{Discussion}\label{ch6/sec:final}

In this paper we have proposed a learning scheme for coherent states of light.
We have presented it in the context of a quantum-enhanced readout of classically-stored binary information, following a recent research line initiated in~\cite{Pirandola2011}.
The reading of information, encoded in the state of a signal that comes reflected by a memory cell, is achieved by measuring the signal and deciding its state to be either the vacuum state or some coherent state of \emph{unknown} amplitude. The effect of this uncertainty is palliated by supplying a large number of auxiliary modes in the same coherent state. We have presented two strategies that make different uses of this (quantum) side information to determine the state of the signal: a collective strategy, consisting in measuring all modes at once and making the binary decision, and a local (E\&D) strategy, based on first estimating---\emph{learning}---the unknown amplitude, then using the acquired knowledge to tune a discrimination measurement over the signal. We have showed that the former outperforms any E\&D strategy that uses a Gaussian estimation measurement over the auxiliary modes. Furthermore, we conjecture that this is indeed the case for \emph{any} (even possibly non-Gaussian) local strategy, based on evidence obtained within a simplified version of the original setting that allowed us to consider completely general measurements.

Previous works on quantum reading rely on the use of specific preparations of nonclassical---namely, entangled---states of light to improve the reading performance of a classical memory \cite{Pirandola2011,Nair2011,Spedalieri2012,Tej2013}. Our results indicate that, when there exists some uncertainty in the states produced by the source (and, consequently, the possibility of preparing a specific entangled signal state is highly diminished), alternative quantum resources---namely, collective measurements---still enhance the reading of classical information using uncorrelated, classical coherent light.
It is worth mentioning that there are precedents of quantum phenomena of this sort providing enhancements for statistical problems involving coherent states. As an example, in the context of estimation of product coherent states, the optimal measure-and-prepare strategy on identical copies of $\ket{\alpha}$ can be achieved by local operations and classical communication (according to the fidelity criterion), but bipartite product states $\ket{\alpha}\!\ket{\alpha^*}$ require entangled measurements~\cite{Niset2007}.

On a final note, the quantum enhancement found here is relevant in the regime of low-energy signals\footnote{Note that here we have only considered sending a single-mode signal. However, in what coherent states are concerned, increasing the number of modes of the signal and increasing the energy of a single mode are equivalent operations.} (small coherent amplitudes). This is in accordance to the advantage regime provided by nonclassical light sources, as discussed in other works \cite{Pirandola2011,Spedalieri2012,RagySciRep}. A low energy readout of memories is, in fact, of very practical interest. While, mathematically, the success probability of any readout protocol could be arbitrarily increased by sending signals with diverging energy, there are many situations where this is highly discouraged. For instance, the readout of photosensitive organic memories requires a high level of control over the amount of energy irradiated per cell. In those situations, the use of signals with very low energy benefits from quantum-enhanced performance, whereas highly energetic classical light could easily damage the memory.

\section*{Acknowledgments}

We acknowledge John Calsamiglia, Ramon Mu\~noz-Tapia and Stefano Pirandola for fruitful discussions and feedback. G.S. was supported by the Spanish MINECO, through Contract No. FIS2008-01236 and FPI Grant No. BES-2009-028117. M.G. was supported by the EPSRC Grant No. EP/J009776/1. G.A. was supported by the Foundational Questions Institute Grant No. FQXi-RFP3-1317.

\appendix
\section{Trace norm for the collective strategy}\label{appD/sec:tracenormcol}

The global states that need to be discriminated in the collective strategy are $\bar{\sigma}_1$ and $\bar{\sigma}_2$. As shown in the main text, the first can be expressed as [cf. Eq.~\eqref{ch6/barsigma1exp}]
\begin{equation}\label{appD/barsigma1exp_bis}
\bar{\sigma}_1 = \sum_{k=0}^\infty c_k [k] \otimes [-\alpha_0] \,,
\end{equation}
whereas the second admits an asymptotic expansion [cf. Eq.~\eqref{ch6/barsigma2exp}]
\begin{equation}\label{appD/barsigma2exp_bis}
\bar{\sigma}_2 \simeq \bar{\sigma}_2^{(0)} + \frac{1}{\sqrt{n}}\bar{\sigma}_2^{(1)} + \frac{1}{n}\bar{\sigma}_2^{(2)}\,,
\end{equation}
as the result of taking the limit $n\to\infty$ up to order $1/n$ in Eq.~\eqref{ch6/barsigma2}. Computing the arising averages (see 
~\ref{ch6/sec:gaussian integrals}), the terms in Eq.~\eqref{appD/barsigma2exp_bis} take the explicit form
%
%
\begin{eqnarray}
\bar{\sigma}_2^{(0)} &=& \sum_{k=0}^\infty   c_k  \ketbrad{k} \otimes \ketbrad{0}  \,, \label{ch6/sigma10} \\
\bar{\sigma}_2^{(1)} &=& \sum_{k=0}^\infty  d_{k+1} \ketbra{k}{k+1} \otimes \ketbra{1}{0} + \tilde{d}_{k-1} \ketbra{k}{k-1} \otimes \ketbra{0}{1} \,, \nonumber \\
& & \label{ch6/sigma11} \\
\bar{\sigma}_2^{(2)} &=& \sum_{k=0}^\infty  e_{k} \ketbrad{k} \otimes \left( \ketbrad{1}-\ketbrad{0}\right) \nonumber \\
&&\,+\, f_{k+2}  \ketbra{k}{k+2} \otimes \ketbra{2}{0} + \tilde{f}_{k-2} \ketbra{k}{k-2} \otimes \ketbra{0}{2}\,, \nonumber \\
& &  \label{ch6/sigma12}
\end{eqnarray}
where
\begin{eqnarray*}
d_{k+1}&=& c_{k+1}\sqrt{k+1} \,,\quad \tilde{d}_{k-1}=c_k \sqrt{k} \,,\\
e_k &=&  c_{k+1}(k+1)  \,,\\
f_{k+2} &=& \frac{1}{\sqrt{2}}c_{k+2}\sqrt{(k+2)(k+1)} \,,\quad
\tilde{f}_{k-2} = \frac{1}{\sqrt{2}}c_k \sqrt{k(k-1)} \,.
\end{eqnarray*}

We now apply perturbation theory to compute the trace norm $\trnorm{\bar{\sigma}_1-\bar{\sigma}_2}$ in the asymptotic limit $n\to\infty$, up to order $1/n$, using Eqs.~\eqref{appD/barsigma1exp_bis} and \eqref{appD/barsigma2exp_bis}. We start by expressing the trace norm as
%
\begin{equation}\label{ch6/tracenorm col}
\trnorm{\bar{\sigma}_1-\bar{\sigma}_2} \simeq\, \trnorm{A+B/\sqrt{n} + C/n \equiv \Gamma} = \sum_j |\gamma_j| \,,
\end{equation}
where $A=\bar{\sigma}_1 -\bar{\sigma}_2^{(0)}$, $B=-\bar{\sigma}_2^{(1)}$, $C=-\bar{\sigma}_2^{(2)}$, and $\gamma_j$ is the $j$th eigenvalue of $\Gamma$, which admits an expansion of the type $\gamma_j = \gamma_j^{(0)} + \gamma_j^{(1)}/\sqrt{n} + \gamma_j^{(2)}/n$.
The matrix $\Gamma$ belongs to the Hilbert space $\mathcal{H}_\infty \otimes \mathcal{H}_3$, i.e., the first mode is described by the infinite dimensional space generated by the Fock basis, and the second mode by the three-dimensional space spanned by the linearly independent vectors $\{\ket{-\alpha_0},\ket{0},\ket{1}\}$ (we will see that the contribution of $\ket{2}$ vanishes, hence it is not necessary to consider a fourth dimension).
Writing the eigenvalue equation associated to $\gamma_j$ and separating the expansion orders, we obtain the set of equations
\begin{align}
&A \psi_j^{(0)} = \gamma_j^{(0)} \psi_j^{(0)} \,, \label{ch6/pert1}\\
&A \psi_j^{(1)} + B \psi_j^{(0)} = \gamma_j^{(0)} \psi_j^{(1)} +\gamma_j^{(1)} \psi_j^{(0)} \,, \label{ch6/pert2}\\
&A \psi_j^{(2)} + B \psi_j^{(1)} + C \psi_j^{(0)} = \gamma_j^{(0)} \psi_j^{(2)} +\gamma_j^{(1)} \psi_j^{(1)} +\gamma_j^{(2)} \psi_j^{(0)} \,, \label{ch6/pert3}
\end{align}
where $\psi_j$ is the eigenvector associated to $\gamma_j$, which also admits the expansion $\psi_j=\psi_j^{(0)}+\psi_j^{(1)}/\sqrt{n}+\psi_j^{(2)}/n$. Eq.~\eqref{ch6/pert1} tells us that $\gamma_j^{(0)}$ is an eigenvalue of $A$ with associated eigenvector $\psi_j^{(0)}$. We multiply \eqref{ch6/pert2} and \eqref{ch6/pert3} by $\bra{\psi_j^{(0)}}$ to obtain
\begin{eqnarray}
\gamma_j^{(1)} &=& \bra{\psi_j^{(0)}} B \ket{\psi_j^{(0)}} \,, \label{ch6/lambda1} \\
\gamma_j^{(2)} &=& \bra{\psi_j^{(0)}} C \ket{\psi_j^{(0)}} + \sum_{l\neq j} \frac{\left|\! \bra{\psi_j^{(0)}} B \ket{\psi_l^{(0)}} \!\right|^2}{\gamma_j^{(0)}-\gamma_l^{(0)}} \,. \label{ch6/lambda2}
\end{eqnarray}
Note that Eq.~\eqref{ch6/lambda2} assumes that there is no degeneracy in the spectrum of $\Gamma$ at zero order (as we will see, this is indeed the case). From the structure of $A$ we can deduce that the form of its eigenvector $\psi_j^{(0)}$ is
\begin{equation}\label{ch6/psi0}
\ket{\psi_{i,\varepsilon}^{(0)}} = \ket{i} \otimes \ket{v_\varepsilon} \,,
\end{equation}
where we have replaced the index $j$ by the pair of indices $i,\varepsilon$. The index $i$ represents the Fock state $\ket{i}$ in the first mode, and the vectors $\ket{v_\varepsilon}$ are eigenvectors of $[-\alpha_0]-[0]$ and form a basis of $\mathcal{H}_3$ in the second mode.
Every eigenvalue of $\Gamma$ is now labelled by the pair of indices $i,\varepsilon$, where $i=0,\ldots,\infty$ and $\varepsilon=+,-,0$: the second mode in $A$ has a positive, a negative, and a zero eigenvalue, to which we associate eigenvectors $\ket{v_+}$, $\ket{v_-}$ and $\ket{v_0}$, respectively. It is straightforward to see that the first two are
\begin{equation}\label{ch6/eigvecs}
\ket{v_\pm} = \frac{1}{2} \left( \frac{\ket{-\alpha_0}+\ket{0}}{N_+} \pm \frac{\ket{-\alpha_0}-\ket{0}}{N_-} \right) \,,
\end{equation}
where $N_\pm = \sqrt{1 \pm e^{-|\alpha_0|^2/2}}$. The zero-order eigenvalues of $\Gamma$ with $\varepsilon=\pm$ are
\begin{equation}\label{ch6/lambda0pm}
\gamma_{i,\pm}^{(0)} = \pm c_i \sqrt{1-e^{-|\alpha_0|^2}} \,.
\end{equation}
The third eigenvector $\ket{v_0}$ is orthogonal to the subspace spanned by $\ket{-\alpha_0}$ and $\ket{0}$, and corresponds to the eigenvalue $\gamma_{i,0}^{(0)} = 0$ \footnote{Note that the zero-order eigenvalues $\gamma_{i,\varepsilon}^{(0)}$ are nondegenerate, hence Eq.~\eqref{ch6/lambda2} presents no divergence problems.}.
This eigenvector only plays a role through the overlap $\braket{1}{v_0}$, which arises in Eqs.~\eqref{ch6/lambda1} and \eqref{ch6/lambda2}. We thus do not need its explicit form, but it will suffice to express $\braket{1}{v_0}$ in terms of known overlaps.

From Eqs.~\eqref{ch6/lambda1} and \eqref{ch6/psi0} we readily see that
$
\gamma_{i,\varepsilon}^{(1)}=0
$.
Using Eqs.~\eqref{ch6/sigma11}, \eqref{ch6/sigma12}, \eqref{ch6/lambda2} and \eqref{ch6/psi0} we can express $\gamma_{i,\varepsilon}^{(2)}$ as
%
%
\begin{eqnarray}
\gamma_{i,\pm}^{(2)} &=&
e_i \left( |\!\braket{0}{v_\pm}\!|^2 - |\!\braket{1}{v_\pm}\!|^2 \right) \nonumber \\
&+& \sum_{\varepsilon} \frac{d_i^2 |\!\braket{0}{v_\pm}\!|^2 |\!\braket{1}{v_\varepsilon}\!|^2}{\gamma_{i,\pm}^{(0)}-\gamma_{i-1,\varepsilon}^{(0)}} + \frac{\tilde{d}_i^2 |\!\braket{1}{v_\pm}\!|^2 |\!\braket{0}{v_\varepsilon}\!|^2}{\gamma_{i,\pm}^{(0)}-\gamma_{i+1,\varepsilon}^{(0)}} \,, \label{ch6/lambda2ov} \\
\gamma_{i,0}^{(2)} &=& 0 \,,\nonumber
\end{eqnarray}
where we have used that, by definition, $\braket{0}{v_0}=\braket{\alpha_0}{v_0}=0$. The overlaps in \eqref{ch6/lambda2ov} are
\begin{eqnarray}
|\!\braket{0}{v_\pm}\!|^2 &=& \frac{1}{2} \left( 1 \mp \sqrt{1-e^{-|\alpha_0|^2}} \right) \,, \label{ch6/ov1}\\
|\!\braket{1}{v_\pm}\!|^2 &=& \frac{|\alpha_0|^2}{2} \frac{1 \pm \sqrt{1-e^{-|\alpha_0|^2}}}{e^{|\alpha_0|^2}-1} \,,\label{ch6/ov2}\\
|\!\braket{1}{v_0}\!|^2 &=& 1-\frac{|\!\braket{1}{-\alpha_0}\!|^2}{1-|\!\braket{0}{-\alpha_0}\!|^2} = 1-\frac{|\alpha_0|^2 e^{-|\alpha_0|^2}}{1-e^{-|\alpha_0|^2}} \,.\label{ch6/ov3}
\end{eqnarray}

Now that we have computed the eigenvalues of $\Gamma$, we are finally in condition to evaluate the sum in the right-hand side of Eq.~\eqref{ch6/tracenorm col}. Incorporating the relevant eigenvalues, given by Eqs.~\eqref{ch6/lambda0pm} and \eqref{ch6/lambda2ov}, it reads
\begin{eqnarray*}
\trnorm{\Gamma} &=& \sum_{i,\varepsilon} \left|\gamma_{i,\varepsilon}^{(0)}+\gamma_{i,\varepsilon}^{(2)}/n\right| \nonumber\\
&=& \sum_{i=0}^\infty \gamma_{i,+}^{(0)} + \frac{1}{n} \gamma_{i,+}^{(2)} - \gamma_{i,-}^{(0)} - \frac{1}{n} \gamma_{i,-}^{(2)} \nonumber\\
&=& \Lambda_+^{(0)} - \Lambda_-^{(0)} + \frac{1}{n} \left( \Lambda_+^{(2)}-\Lambda_-^{(2)} \right) \,,
\end{eqnarray*}
where
\begin{equation*}
\Lambda_\pm^{(0)} = \sum_{i=0}^\infty \gamma_{i,\pm}^{(0)} = \pm \sqrt{1-e^{-|\alpha_0|^2}}
\end{equation*}
(recall that $\sum_{i=0}^\infty c_i = 1$), and
\begin{equation}\label{appD/Lambda2pm}
\Lambda_{\pm}^{(2)}=\sum_{i=0}^\infty \gamma_{i,\pm}^{(2)} = \pm \frac{\mu^2 e^{-|\alpha_0|^2/2}}{2\sqrt{e^{|\alpha_0|^2}-1}} \left( 1-\frac{\mu^2+1}{2\mu^2+1} \frac{|\alpha_0|^2 \left(2e^{|\alpha_0|^2}-1\right)}{e^{|\alpha_0|^2}-1} \right) \,.
\end{equation}
\\

\section{\boldmath Conditional probability $p(\beta|\alpha)$, Eq.~\eqref{ch6/heterodyne_prob_ab}}\label{appD/sec:heterodyne_prob}

Given two arbitrary Gaussian states $\rho_A, \rho_B$, the trace of their product is
\begin{equation}\label{ch6/traceAB}
\tr (\rho_A \rho_B) = \frac{2}{\sqrt{\det(V_A+V_B)}} e^{-\delta^T (V_A+V_B)^{-1} \delta} \,,
\end{equation}
where $V_A$ and $V_B$ are their covariance matrices and $\delta$ is the difference of their displacement vectors. For the states $\rho_A \equiv [\sqrt{n}\alpha]$ and $\rho_B \equiv E_{\bar{\beta}}$, we have
\begin{eqnarray*}
V_A &=& \begin{pmatrix} 1 & 0 \\ 0 & 1 \end{pmatrix} \,,\quad
V_B = R \begin{pmatrix} e^{-2r} & 0 \\ 0 & e^{2r} \end{pmatrix} R^T \,, \\
R &=& \begin{pmatrix} \cos \phi & -\sin \phi \\ \sin \phi & \cos \phi \end{pmatrix} \,, \\
\delta &=& (\sqrt{n} a_1-\bar{b}_1,\sqrt{n} a_2-\bar{b}_2) \,,
\end{eqnarray*}
where $\alpha = a_1 + i a_2$, $\bar{\beta} = \bar{b}_1 + i \bar{b}_2$, $r$ is the squeezing parameter, and $\phi$ indicates the direction of squeezing in the phase space. In terms of $\alpha$ and $\bar{\beta}$, Eq.~\eqref{ch6/traceAB} reads
\begin{equation*}
\tr (\rho_A \rho_B) = \frac{1}{\pi \cosh r} e^{-|\sqrt{n} \alpha - \bar{\beta}|^2-{\rm Re}[(\sqrt{n} \alpha-\bar{\beta})^2 e^{-i 2 \phi}] \tanh r} \,.
\end{equation*}

\section{Trace norm for the E\&D strategy}\label{ch6/sec:eyd_tracenorm}

For assessing the performance of the E\&D strategy, we want to obtain the error probability in discriminating the state $[0]$ and the posterior state $\rho(\beta)$, resulting from a heterodyne estimation of the state of the auxiliary mode that provides the estimate $\beta$.
Under a local Gaussian model around $\alpha_0$ parametrised by the complex variables $u$ and $v$, these states transform into $[-\alpha_0]$ and $\rho(v)$, respectively, where the second is given by
\begin{equation*}
\rho(v)=\int p(u|v) \, [u/\sqrt{n}] \, d^2u \,,
\end{equation*}
and where $p(u|v)$ is given by Eq.~\eqref{ch6/puv}. The error probability is determined by the trace norm $\trnorm{[-\alpha_0]-\rho(v)}$ [cf. Eq.~\eqref{ch6/tracenorm_eyd}]. To compute it, we first series expand $\rho(v)$ in the limit $n\to\infty$, up to order $1/n$.
We name the appearing integrals of $u,u^*,|u|^2,u^2$, and $(u^*)^2$ over the probability distribution $p(u|v)$ as $I_1,I_1^*,I_2,I_3$, and $I_3^*$, respectively. This allows us to write the trace norm 
as
\begin{equation*}
\trnorm{[-\alpha_0]-\rho(v)} \simeq \trnorm{A'+B'/\sqrt{n}+C'/n \equiv \Phi} = \sum_\kappa |\lambda_\kappa| \,,
\end{equation*}
where
\begin{eqnarray*}
A' &=& \ketbrad{-\alpha_0} - \ketbrad{0} \,,\\
B' &=& -I_1 \ketbra{1}{0} - I_1^* \ketbra{0}{1} \,,\\
C' &=& -I_2 \left(\ketbrad{1}-\ketbrad{0}\right) - \frac{1}{\sqrt{2}}\left(I_3\ketbra{2}{0}+I_3^*\ketbra{0}{2}\right) \,,
\end{eqnarray*}
and $\lambda_\kappa$ is the $\kappa$th eigenvalue of $\Phi$, which admits the perturbative expansion $\lambda_\kappa = \lambda_{\kappa}^{(0)}+\lambda_{\kappa}^{(1)}/\sqrt{n}+\lambda_{\kappa}^{(2)}/n$, just as its associated eigenvector
$\varphi_\kappa = \varphi_{\kappa}^{(0)} + \varphi_{\kappa}^{(1)}/\sqrt{n} + \varphi_{\kappa}^{(2)}/n$.
Up to order $1/n$, the matrix $\Phi$ has effective dimension 4 since it belongs to the space spanned by the set of linearly independent vectors $\{\ket{-\alpha_0},\ket{0},\ket{1},\ket{2}\}$. Hence the index $\kappa$ has in this case four possible values, i.e., $\kappa=+,-,3,4$. The zero-order eigenvalues $\lambda_\kappa^{(0)}$, which correspond to the eigenvalues of the rank-2 matrix $A'$, are
\begin{equation*}
\lambda_\pm^{(0)} = \pm \sqrt{1-e^{-\alpha_0^2}} \,,\quad \lambda_{3}^{(0)}=\lambda_{4}^{(0)} = 0 \,
\end{equation*}
(recall that $\alpha_0\in\mathbb{R}$). Their associated eigenvectors are $|\varphi_{\kappa}^{(0)}\rangle = \ket{v_\kappa}$, where $\ket{v_\pm}$ is given by Eq.~\eqref{ch6/eigvecs}, and, by definition, $\braket{v_\kappa}{-\alpha_0}=\braket{v_\kappa}{0}=0$ for $\kappa=3,4$. From analogous expressions to Eqs.~\eqref{ch6/lambda1} and \eqref{ch6/lambda2} we can write the first and second-order eigenvalues as
\begin{eqnarray*}
\lambda_{\kappa}^{(1)} &=& -I_1 \braket{v_\kappa}{1}\braket{0}{v_\kappa} - I_1^* \braket{v_\kappa}{0}\braket{1}{v_\kappa} \,,\\
\lambda_{\kappa}^{(2)} &=& I_2 \left(|\!\braket{v_\kappa}{0}\!|^2\!\! -\! |\!\braket{v_\kappa}{1}\!|^2 \right) -\!\frac{1}{\sqrt{2}} \left( I_3 \braket{v_{\kappa}}{2}\braket{0}{v_{\kappa}} + I_3^* \braket{v_{\kappa}}{0}\braket{2}{v_{\kappa}}\right) \nonumber\\
& + &  \sum_{\xi \neq \kappa}
\left(  |I_1|^2 \frac{ |\!\braket{v_\xi}{1}\!|^2 |\!\braket{v_{\kappa}}{0}\!|^2 + |\!\braket{v_\xi}{0}\!|^2 |\!\braket{v_{\kappa}}{1}\!|^2 }{\lambda_{\kappa}^{(0)}-\lambda_{\xi}^{(0)}} \right.\nonumber\\
&+& \left.\frac{I_1^2 \braket{v_\xi}{1}\braket{v_{\kappa}}{1}\braket{0}{v_{\kappa}}\braket{0}{v_\xi} + (I_1^*)^2 \braket{1}{v_\xi}\braket{1}{v_{\kappa}}\braket{v_{\kappa}}{0}\braket{v_\xi}{0} }{\lambda_{\kappa}^{(0)}-\lambda_{\xi}^{(0)}}
\right).
\end{eqnarray*}
The needed overlaps for computing $\lambda_{\kappa}^{(1)}$ and $\lambda_{\kappa}^{(2)}$ are given by Eqs.~\eqref{ch6/ov1}, \eqref{ch6/ov2}, and
\begin{align}
\braket{v_\pm}{0} &= \frac{1}{2} \left( N_+ \mp N_- \right) \,,\nonumber\\
\braket{v_\pm}{1} &= \frac{1}{2} (-\alpha_0) e^{-\alpha_0^2/2} \left( \frac{1}{N_+} \pm \frac{1}{N_-} \right) \,,\nonumber\\
|\!\braket{v_3}{1}\!|^2 &= 1-\frac{|\!\braket{1}{-\alpha_0}\!|^2}{1-|\!\braket{0}{-\alpha_0}\!|^2-|\!\braket{2}{-\alpha_0}\!|^2} \,, \label{ch6/ov_v3_1}\\
|\!\braket{v_4}{1}\!|^2 &= \frac{|\!\braket{1}{-\alpha_0}\!|^2 |\!\braket{2}{-\alpha_0}\!|^2}{\left(1-|\!\braket{0}{-\alpha_0}\!|^2\right)\left(1-|\!\braket{0}{-\alpha_0}\!|^2-|\!\braket{2}{-\alpha_0}\!|^2\right)} \label{ch6/ov_v3_2}\,.
\end{align}
The expressions for the overlaps \eqref{ch6/ov_v3_1} and \eqref{ch6/ov_v3_2} actually depend on the dimension of the space that we are considering (four in this case), and they are not unique: there are infinitely many possible orientations of the orthogonal pair of vectors $\{\ket{v_3},\ket{v_4}\}$ such that both of them are orthogonal to the plane formed by $\{\ket{-\alpha_0},\ket{0}\}$, which is the only requirement we have. However, one can verify that this choice does not affect the trace norm $\trnorm{\Phi}$, thus we are free to choose the particular orientation that, in addition, verifies $\braket{v_3}{2}=0$, yielding the simple expressions \eqref{ch6/ov_v3_1} and \eqref{ch6/ov_v3_2}.

Finally, we write down the trace norm as
\begin{eqnarray}
\trnorm{\Phi} &=& \sum_\kappa |\lambda_{\kappa}^{(0)} + \lambda_{\kappa}^{(1)}/\sqrt{n} + \lambda_{\kappa}^{(2)}/n| \nonumber\\
&=& \lambda_+^{(0)} - \lambda_-^{(0)} + \frac{1}{\sqrt{n}} \left(\lambda_+^{(1)}-\lambda_-^{(1)}\right) \nonumber\\
&& +\; \frac{1}{n} \left(\lambda_+^{(2)}-\lambda_-^{(2)}+|\lambda_3^{(2)}|+|\lambda_4^{(2)}|\right)\,,\label{ch6/tracenorm_eyd_2}
\end{eqnarray}
which we use now to obtain the asymptotic expression for the average error probability, defined in Eq.~\eqref{ch6/perror_eyd}. Recall Eq.~\eqref{ch6/pv} and note that we have to average Eq.~\eqref{ch6/tracenorm_eyd_2} over the probability distribution $p(v)$. Regarding this average, it is worth taking into account the following considerations. First, the $v$-dependence of the eigenvalues comes from $I_1,I_2,I_3$, and its complex conjugates. The integrals needed are given in the last part of 
\ref{ch6/sec:gaussian integrals}. Second, 
the integration yields
$\lambda_{\kappa}^{(1)}=0$ and hence the order $1/\sqrt{n}$ term vanishes, as it should. And third, the second-order eigenvalues $\lambda_3^{(2)}$ and $\lambda_4^{(2)}$ are $v$-independent and positive, so we can ignore the absolute values in Eq.~\eqref{ch6/tracenorm_eyd_2}.
Putting all together, we can express the asymptotic average error probability of the E\&D strategy as
\begin{equation}\label{appD/perror_eyd_final}
P_{\rm e}^{\rm E\&D} \equiv P_{\rm e}^{\rm E\&D}(n\to\infty) \simeq \frac{1}{2}\left(1-\sqrt{1-e^{-\alpha_0^2}} + \frac{1}{n} \Delta^{\rm E\&D} \right)\,,
\end{equation}
where
\begin{equation}\label{appD/DeltaEyD}
\Delta^{\rm E\&D} = -\frac{1}{2} \left[\lambda_3^{(2)}+\lambda_4^{(2)} + \int p(v) \left(\lambda_+^{(2)}-\lambda_-^{(2)}\right) dv \right] \,.
\end{equation}

Making use of Eqs.~\eqref{appD/perror_eyd_final} and \eqref{ch6/perror known} we can readily compute the excess risk of the E\&D strategy:

\begin{align}\label{appD/excessrisk_eyd_r}
R^{\rm E\&D}(r) &= n \lim_{\mu\to\infty} \left(P_{\rm e}^{\rm E\&D}-P_{\rm e}^*\right) \nonumber\\
&\!\!\!\!\!\!= \frac{e^{-\alpha_0^2}}
{
16 \sqrt{1-e^{-\alpha_0^2}} \left(e^{\alpha_0^2}-1\right)
}
\left\{  \left[ 4e^{\alpha_0^2} \left(1-e^{\alpha_0^2}\right)\left(\sqrt{1-e^{-\alpha_0^2}}-1\right) \right.\right.\nonumber\\
&\!\!\!\!\!\! \quad\left.\left. +\, \alpha_0^2 \left(4e^{\alpha_0^2}\sqrt{1-e^{-\alpha_0^2}}-2\right) \right] \cosh^2r + \alpha_0^2\sinh(2r)  \right\} \,.
\end{align}
\\

\section{Gaussian integrals}\label{ch6/sec:gaussian integrals}

At many points in this paper, we integrate complex-valued functions over the complex plane, weighted by the bidimensional Gaussian probability distribution $G(u)$. This section gathers the integrals that we need. Recall that $G(u)$ is defined as
\begin{equation*}
G(u) = \frac{1}{\pi \mu^2} e^{-|u|^2/\mu^2} \,,\quad u\in\mathbb{C} \,.
\end{equation*}
Expressing $u$ either in polar or Cartesian coordinates in the complex plane, i.e., $u = r e^{i\theta} = u_1+i u_2$, one can readily check that $G(u)$ is normalised:
\begin{eqnarray*}
\int G(u) d^2u &=& \int_0^\infty \int_0^{2\pi} \frac{1}{\pi \mu^2} e^{-r^2/\mu^2} r dr d\theta=1 \,,\\
\int G(u) d^2u &=& \int_{-\infty}^{\infty} \int_{-\infty}^{\infty} \frac{1}{\pi \mu^2} e^{(-u_1^2-u_2^2)/\mu^2} du_1 du_2 = 1 \,.
\end{eqnarray*}
The average of a coherent state $[u]$ over the probability distribution $G(u)$ can be computed by expressing $\ket{u}$ in the Fock basis $\{\ket{k}\}$. It gives
\begin{equation*}\label{appD/int_coherent}
\int G(u) [u] d^2u = \sum_{k=0}^\infty c_k [k] \,,\quad c_k=\frac{\mu^{2k}}{(\mu^2+1)^{k+1}} \,.
\end{equation*}
Note that the result of averaging a coherent state over $G(u)$ is nothing else than a thermal state with average photon number $\mu^2$.

Variations of the previous integral 
with different complex functions that we use are
%
%
\begin{eqnarray*}
\int G(u) u [u] d^2u &=& \sum_{k=0}^\infty c_{k+1} \sqrt{k+1} \ketbra{k}{k+1} \,, \\
\int G(u) u^* [u] d^2u &=& \sum_{k=0}^\infty c_k \sqrt{k} \ketbra{k}{k-1} \,, \\
\int G(u) |u|^2 [u] d^2u &=& \sum_{k=0}^\infty c_{k+1}  (k+1) \ketbrad{k} \,, \\
\int G(u) u^2 [u] d^2u &=& \sum_{k=0}^\infty c_{k+2} \sqrt{k+2}\sqrt{k+1} \ketbra{k}{k+2} \,, \\
\int G(u) \left(u^*\right)^2 [u] d^2u &=& \sum_{k=0}^\infty c_k \sqrt{k}\sqrt{k-1} \ketbra{k}{k-2} \,,
\end{eqnarray*}
and
\begin{eqnarray*}
\int G(u) (u+u^*) d^2u &=& 0 \label{ch6/int u1}\,,\\
\int G(u) (u+u^*)^2 d^2u &=& 2\mu^2 \label{ch6/int u2}\,,\\
\int G(u) |u|^2 d^2u &=& \mu^2 \label{ch6/int u3}\,.
\end{eqnarray*}

%

For the computations in 
\ref{ch6/sec:eyd_tracenorm} we also need to perform Gaussian integrals, this time over the probability distribution $p(v)$, defined in Eq.~\eqref{ch6/pv}. We make use of
\begin{eqnarray*}
\int p(v) I_1 d^2v &=& \int p(v) I_1^* d^2v = 0 \,,\label{appD/intI1}\\
\int p(v) I_3 d^2v &=& \int p(v) I_3^* d^2v = 0 \,,\label{appD/intI3}\\
\int p(v) I_2 d^2v &=& \mu^2 \,,\nonumber\\
\int p(v) I_1^2 d^2v &=& \int p(v) (I_1^*)^2 d^2v \nonumber\\
&=& \frac{\mu^4 \sinh(2r)}{(2\mu^2+1)\cosh(2r)+2\mu^2(\mu^2+1)+1} \,,\nonumber\\[1em]
\int p(v) |I_1|^2 d^2v &=& \frac{\mu^4 (\cosh(2r)+2\mu^2+1)}{(2\mu^2+1)\cosh(2r)+2\mu^2(\mu^2+1)+1} \,.\nonumber
\end{eqnarray*}


\section*{References}

\bibliographystyle{iopart-num}
\bibliography{learningbib}

\end{document}